# Impulse-Response Approach to Elastobaric Model for Proteins

Robert D. Young

*Department of Physics, Illinois State University*
*Normal, IL 61790-4560*

(Dated January 4, 2018)

## Abstract

A novel energy landscape model (ELM) for proteins recently explained a collection of incoherent, elastic neutron scattering data from proteins. The ELM of proteins considers the elastic response of the proton and its environment to the energy and momentum exchanged with the neutron. In the ELM, the elastic potential energy is expressed as a sum of a temperature dependent term resulting from equipartition of potential energy among the active degrees of freedom and a wave-vector transfer dependent term resulting from the elastic energy stored by the protein during the neutron scattering event. The elastic potential energy involves a new elastobaric coefficient that is proportional to the product of two factors – one factor depends on universal constants and the other on the incident neutron wave-vector per degree of freedom. The ELM was tested for dry protein samples with an elastobaric coefficient corresponding to 3 degrees of freedom. A discussion of the data requirements for additional tests of ELM is presented resulting in a call for published data that have not been preprocessed by temperature and wave-vector dependent normalizations.

## I. Introduction

A protein is a folded chain of amino acids with a covalently bonded structure as well as non-bonded interactions that involve hydration waters and bulk solvent. Protein structure and function have been analyzed extensively [1]. Proteins are also dynamic systems that experience continual fluctuations [2]. Fluctuations, in fact, are essential for protein function. Mössbauer absorption spectroscopy, incoherent neutron scattering, dielectric spectroscopy, and other techniques are used to probe protein fluctuations. However, understanding of protein dynamics is not fully developed.

Incoherent neutron scattering has developed into a mature experimental tool for probing protein fluctuations [3]. It is customary to use models to extract a measure of spatial fluctuations in the form of the mean-square displacement of the protein hydrogen atom that scatters the neutron. This class of models is called SMM for spatial motion models. Recently shortcomings of the SMM have been outlined in two papers by Frauenfelder and collaborators [4,5]. As a replacement for the SMM, these authors developed a model called ELM for energy landscape model in which the elastic response of a protein plays a central role. The ELM model envisions a quite different role for the dynamic momentum (or equivalently, wave-vector) transfer in incoherent neutron scattering. The ELM passed initial tests by describing elastic incoherent neutron scattering data for several dehydrated proteins [5]. The ELM involves two parameters – one called the elastobaric coefficient $\chi$ with dimension kelvin-Ångstom (K-Å) and the other called $\Theta$ with dimension kelvin (K). Here an impulse-response approach of the ELM is developed in which elastobaric coefficient $\chi$ is described by universal constants, the initial neutron wave-vector, and the number of active degrees of freedom, $d$, of the target proton and its environment.

## II. Impulse-Response Approach of Elastobaric Model for Neutron Scattering

The neutron (N) interaction with a proton (P) covalently bound in a molecule results in a time-dependent force $F(t)$ on the proton and its immediate molecular environment [5]. The analytic form of $F(t)$ requires knowledge of the strong interaction between the neutron and proton and, also, the bonded and non-bonded interactions between the proton and its molecular environment. To quantify the elastic response of the protein to the incident neutron without knowing these complicated forces, an impulse-response approach is followed. The total momentum transfer $\delta P$ to the proton and its moiety during the N-P interaction is given by the impulse, $\delta P = \hbar Q = \int F(t) dt$ where the time integral extends over the N-P interaction time, $\delta t$, and $Q$ is the wave-vector transfer. If $\bar{F}$ is the average force on a proton and its moiety over the N-P interaction time, then $\bar{F}\delta t \approx \delta P = \hbar Q$. The average work done by $\bar{F}$, acting over a curvilinear displacement $\delta l$, results in potential energy storage $\delta U(Q)$ in the proton environment where $\delta U(Q) = \bar{F}\delta l = \hbar Q \delta l / \delta t$. The speed $v_o$ of the incident neutron is not changed by elastic scattering so is approximated as $v_o = \delta l / \delta t$. The potential energy stored then is $\delta U(Q) = \hbar Q v_o$. Using the momentum of the incident neutron, the average speed can be estimated by $m_n v_o = \hbar k_o$ where $k_o$ is the central wave vector of the incident neutron wave packet. The potential energy stored in the proton environment is then

$$\delta U(Q) = \hbar Q v_o = \frac{\hbar^2 k_o}{m_n} Q. \tag{1}$$

Some, or all, of the transferred energy might be kinetic during the interaction, but the energy is ultimately stored as potential energy as in Eq. (1).

The potential energy of the proton and its environment for the active degrees of freedom is assumed to be $k_B T/2$ for every relevant degree of freedom. For $d$ active degrees of freedom, the potential energy $U_o$ before the interaction is given by $U_o(T) = d k_B T/2$. The total potential energy $U$ of the proton and its environment for a specific population of proteins that scatter with momentum transfer $Q$ is then given by

$$U(T,Q) = U_0(T) + \delta U(Q) = \frac{d}{2} k_B T + \frac{\hbar^2 k_o}{m_n} Q = \frac{d}{2} k_B (T + \chi Q) \tag{2}$$

where the elastobaric coefficient $\chi$ is given by

$$\chi = \left(\frac{2\hbar^2}{m_n k_B}\right)\frac{k_o}{d}. \tag{3}$$

Eq. (2) can be rearranged to read

$$T = T^* - \chi Q. \tag{4}$$

$T^*$ has dimension K but is a scaled potential energy.

Frauenfelder named the model underlying Eqs. (2) and (3) the elastobaric model [5,6,7,8]. The elastobaric coefficient $\chi$ is proportional to the initial neutron wave-vector per degree of freedom of the target proton and environment, or equivalently, the initial neutron momentum per degree of freedom. The proportionality factor, $2\hbar^2/m_n k_B = 95.7 \text{ K}-\text{Å}^2$, is given in terms of universal constants. Once $\chi$ is determined experimentally, the number of active degrees of freedom of the proton and its environment is known. Eqs. (2) and (3) are the main results for the elastobaric model in the impulse-response approach. As an example, the IN13 neutron spectrometer at Institut Laue-Langevin (ILL) involves scattering with a neutron beam having incident wave-vector $k_o = 2.82 \text{ Å}^{-1}$. According to Eq. (3), $\chi = 269.9/d$ K-Å for the neutron beam at IN13. Fig. 1 shows the variation of $\chi$ with changing degrees of freedom for IN13.

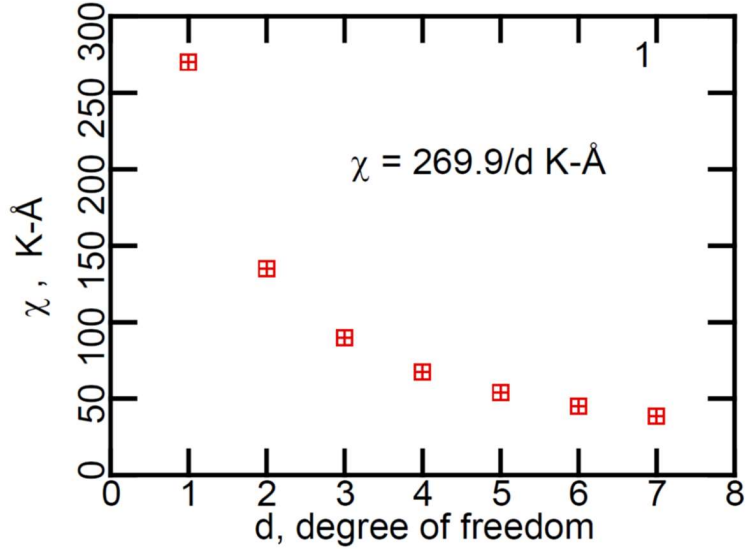

Figure 1. The elastobaric parameter $\chi$ vs degrees of freedom $d$ for $k_o = 2.82 \text{ Å}^{-1}$.

The elastic intensity for incoherent neutron scattering on protein samples can be approximated by

$$S(0,Q,T) \approx 1 - \left(\frac{T^*}{\Theta}\right)^2 = 1 - \left(\frac{T + \chi Q}{\Theta}\right)^2, \tag{5}$$

Eq. (5) is used to fit elastic scattering data for several proteins [5]. For dry protein samples or for samples below about 200 K, the two parameters have values of $\chi \approx 91$ K-Å and $\Theta \approx 750$ K and describe the data well. See, for example, Figs. 5B and 5C in Reference [5]. The elastic incoherent neutron scattering data for the dry protein samples were taken on the IN13 neutron spectrometer so that Eq. (3) gives $\chi \approx 90$ K-Å for dimension $d = 3$, a value for $\chi$ close to the value found in the published fits [5].

### III. Variation in Degrees of Freedom

Protein samples may have a heterogeneity among the individual proteins of the sample in the elastic response to the incident neutron. For example, the number of degrees of freedom involved in the elastic response may differ among the proteins of the sample. For the proteins treated in Reference [5] the number of degrees of freedom was 3 as shown above. But there are differences in neutron elastic scattering for different proteins [3,5,9-12]. The possibility that differing number of degrees of freedom are activated during the neutron scattering process over an ensemble of proteins in a sample is real. This section explores how to quantify this possibility.

Fig. 1 indicates a slowing in the change in $\chi$ as the number of degrees of freedom increases. For example, if the dimension changes from 2 to 3, then $\chi$ changes from 135 K-Å to 90 K-Å. Data for hydrated samples above 200 K indicate that hydration activates new degrees of freedom in hydrated samples [10,11,12]. So, for example, if the degrees of freedom are increased to 6 at high temperature, then $\chi \approx 45$ K-Å. However, if the degrees of freedom changes from 5 to 6, $\chi$ changes from 54 K-Å to 45 K-Å. These observations suggest that low temperatures, below 200 K, and dry protein samples would be best to search for the effects of differences in activated degrees of freedom for different proteins.

To quantify the discussion above, Fig. 2 displays simulations at 140 K for the elastic intensity $S(0, Q, T = 140 \text{ K})$. The parameter $\Theta \approx 750$ K in all simulations and the incident wave-vector $k_o = 2.82$ Å$^{-1}$. The simulations for degrees of freedom $d = 2, 3, 4$ are shown as points at 9 values of wave-vector transfer $Q$ (see caption in Fig. 2). Two additional sets of points are simulated using an assumed sample heterogeneity in which different number of degrees of freedom are affected by the scattering (50-50 % for $d = 3, 4$ and 25-75 % for $d = 2, 3$). The parameter $\Theta \approx 750$ K in all simulations and the incident wave-vector $k_o = 2.82$ Å$^{-1}$.

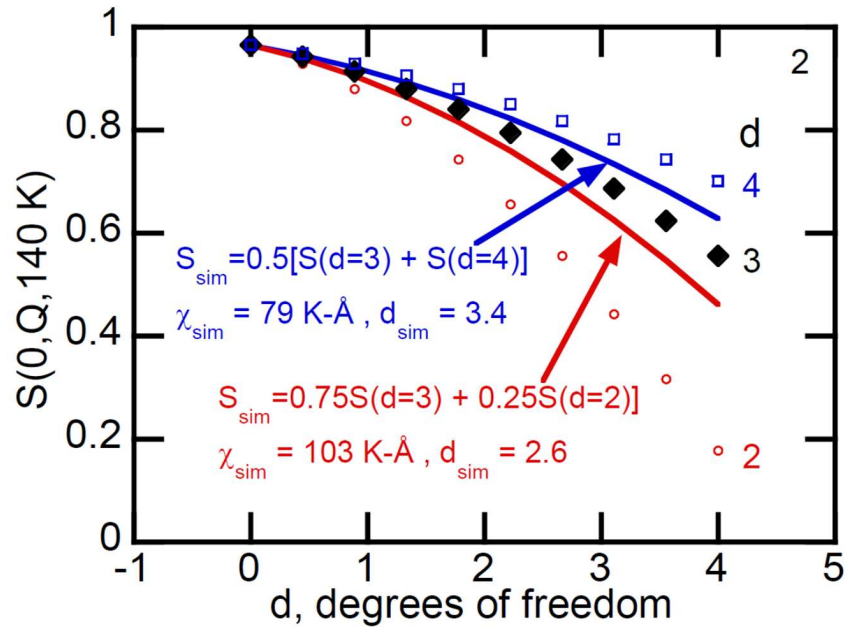

Figure 2. Simulated values of elastic intensity $S(0, Q, 140\ K)$ plotted vs Q for three values of the degrees of freedom ($d = 2$ red open circles, $d = 3$ black closed diamonds, $d = 4$ blue open squares). Fits of Eq. (5) to simulated elastic intensities for the admixtures of 50-50 % $d = 3$ and 4 (blue solid line) and 25-75 % $d = 2$ and 3 (red solid line) are shown. The simulated elastobaric parameters are given together with the degrees of freedom. $k_o = 2.82\ \text{Å}^{-1}$.

Eq. (5) is then used to fit the simulated points with the result summarized in Fig. 2. The two values of $\chi$ and $d$ from the fits to the simulated points are $\chi_{sim} \approx 79\ \text{K-Å}$, $d_{sim} = 3.4$ (50-50 % for $d = 3$, 4) and $\chi_{sim} \approx 103\ \text{K-Å}$, $d_{sim} = 2.6$ (25-75 % for $d = 2$, 3). These simulations do not imply fractal dimensions for proteins. The heterogeneity described above is a result of a heterogeneous sample from the standpoint of possible differing number of degrees of freedom being active during neutron scattering in different proteins

The results in the preceding paragraphs and Fig. 2 indicate that further progress requires additional elastic incoherent neutron scattering data on proteins data that have not preprocessed by temperature and wave-vector dependent "normalizations". Such normalizations distort the primary data obtained in the experiment [5].

## IV. Conclusion

Frauenfelder and collaborators have shown that the wave-vector transfer $Q$ in elastic neutron scattering from protein samples results in an inhomogeneity of the target during passage of the neutron [5]. The resulting energy landscape model (ELM) depends crucially on the elastic response of the protein to the perturbation caused by the neutron-proton interaction. The effect is described by the elastobaric coefficient $\chi$. Here we introduce an impulse-response analysis of the neutron-proton interaction by analyzing the energy and momentum transfer between the neutron and the proton and its surroundings. This model shows that $\chi = \left(2\hbar^2/m_n k_B\right)\left(k_o/d\right)$. The predicted value of $\chi$ agrees closely with the fitted value of $\chi$ in Reference [5]. We also suggest a path to further test the ELM by the publication of primary elastic neutron scattering data that have not been preprocessed by arbitrary temperature and wave-vector dependent "normalizations." Finally, the parameter $\Theta$ requires continued analysis to determine its full meaning [5].

**Acknowledgement**

I thank my colleagues Hans Frauenfelder and Paul Fenimore for continued and deep discussions.